# A lasing mechanism based on absorption boundary of gain materials

Jinwei Shi[1*], Shujing Chen[1], Wenjun Fan[1], Xiangyu Kong[1], Dahe Liu[1,*], Lily Zu[2]

1 Applied Optics Beijing Area Major Laboratory, Department of Physics, Beijing Normal University, Beijing 100875, China.

2 College of Chemistry, Beijing Normal University, Beijing 100875, China.

* Corresponding author: shijinwei@bnu.edu.cn, dhliu@bnu.edu.cn

## ABSTRACT

A new kind of mechanism of lasing is investigated experimentally. It is quite different from the traditional laser with cavity and the random laser with random scattering. In this mechanism, the intensity-dependent refractive index effect and thermal lensing effects of the pump beam induce a large gradient of the refractive index in the gain material, which forms a passive equivalent boundary that provides the feedback in the lasing system. A real lasing system, a liquid disk laser, is performed, it achieves 2-D omnidirectional radiation with a high efficiency of 28%, its radiation spectral property can be explained by resonant Raman scattering.

## 1. Introduction

Up to now, there are two lasing mechanisms: the traditional laser with artificial cavity [1-4] and the random laser with random scattering [5-8]. It is known that, the positive feedback is the key factor in a lasing system. A. L. Schawlow and C. H. Towns proposed the resonator based on Fabry-Perot etalon [1]. R. V. Ambartsumyan and N. G. Basov suggested the disordered structure with random scattering [5]. The artificial cavity in the traditional laser and the random scattering in the disordered structure provide the positive feedback. Besides, some researchers consider the amplified spontaneous emission (ASE) as a kind of lasing although it is different from general lasing process. Is there a new lasing mechanism different from the previous two kinds of feedback? The answer is definitely "yes". Actually, in 1971, C.V.Shank, J.E.Bjorkholm and H. Kogelnik had described a distributed-feedback dye laser (DFL) without mirrors or particle scattering [9]. The feedback is provided by the interference of two incident laser inside the active medium in the direction parallel to the surface.

In this present work, we report a liquid disk laser system [10-18] with high efficiency 2-D omnidirectional radiation, in which there is also no artificial cavity, like mirrors/particles, or distributed-feedback to provide feedback. In the system, the lasing is supported by the feedback of a passive resonator, which is consist of an equivalent boundary induced by intensity-dependent refractive index effect and thermal lensing effects of a pump beam due to the resonant absorption of the gain material. The passive cavity is formed by the pump beam. There will be no cavity, even no any structure, if the pump beam is not applied. The following is our work in details.

## 2. Experiment and experimental results

Figure 1(a) shows schematically the optical layout of the experiments. The pumping laser beam with a flat top transverse profile is vertically incident into the gain material from the top of the solution surface. A Nd: YAG laser running at 532 nm was used as the pump source. Its repetition rate was 10 Hz, and the pulse duration was 10 ns. The gain material used was rhodamine 6G (R6G) dissolved in methanol with the concentration varying from $1C_{R0}$ to $4C_{R0}$, where $C_{R0} = 1.3 \times 10^{-3}$ mol/L. The R6G solution was in a cuvette (10-mm width, 100-mm length and 50-mm height). The photoluminescence (PL) spectrum was detected by a spectrometer (Ocean Optics model Maya Pro 2000 with a spectral resolution of 0.4 nm). Figure 1(b) shows the real system. It can be seen that: 1) the system can excite a strong directional radiation although there is no traditional cavity, and 2) the beam of this radiation is almost parallel to the surface of the gain material.

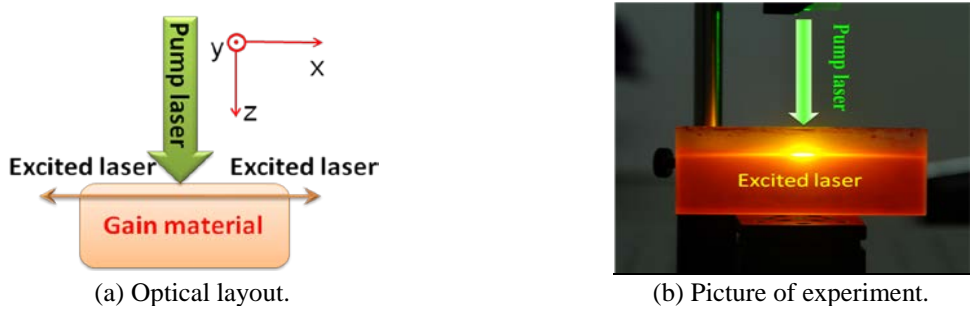

(a) Optical layout.    (b) Picture of experiment.
Fig. 1. Set-up geometry and experimental result.

Figure 2 shows the typical PL spectra of the directional radiation under varying pump power. When the pump power is lower than a certain threshold, the spectrum is a wide PL profile; however, when the pump power is higher than this threshold, several narrow peaks appear at the top of the PL profile. If the pump power is raised further, these peaks become higher. The line widths of the narrow peaks are around 0.5 nm, which is close to the resolution limit of the spectrometer used. The threshold power density of the narrow peak is about 13.93 MW/cm$^2$, which is considerably low. These features suggested that this directional radiation is not common fluorescence but laser. It should be noted that the feedback by the boundary of the cuvette can be ignored, because it has no effect on the lasing result when we change the position or shape of the cuvette boundary and the pumping laser beam. On the other hand, the possible frequencies of the peaks do not change with the shape and intensity of the pump beam as long as lasing is launched. Only the pump wavelength can change the frequencies of the peaks. This phenomenon will be discussed later.

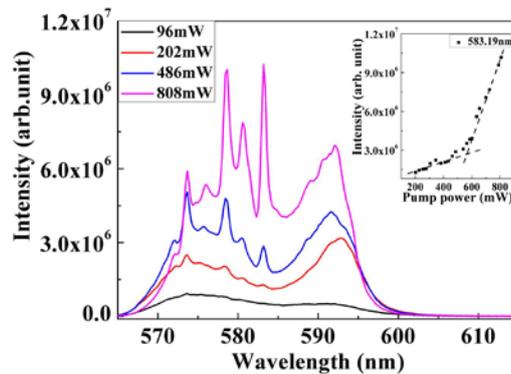

Fig. 2. PL spectra under three different pump powers at 532 nm. The wide peak around 590 nm is from re-emission, and can be understood later by the result shown in Fig. 6. The inset is the measured result under a different pump power, which shows the threshold effect of the emission at 583.19 nm.

There is no traditional cavity in the laser system shown in Fig. 1. Additionally, it is not a random laser or a DFL, meaning that there must be a new lasing mechanism for such a system. The following experiments were carried out further to reveal the new mechanism.

First, the cross section of the pump beam was chosen as a circle, square and rectangle. It is found that both the spectral profile (the envelope) and intensity of the radiation are dependent on the cross section of the pump beam. Along the long side of the cross section, the PL is stronger and the linewidth of the peaks are narrower. This phenomenon suggests that an equivalent boundary exists in the pump area of the gain material and provides the feedback. The strong absorption of the pumping laser (532 nm) by the R6G solution will induce intensity-dependent refractive index and thermal lensing effects [19], and cause a large gradient in the refractive index of the gain material from the center to the outer region. This results in an equivalent boundary, i.e. an equivalent cavity, in the transverse plane, which provides strong feedback and keeps the lasing inside rather than vertical to the transverse plane. It will be discussed again later.

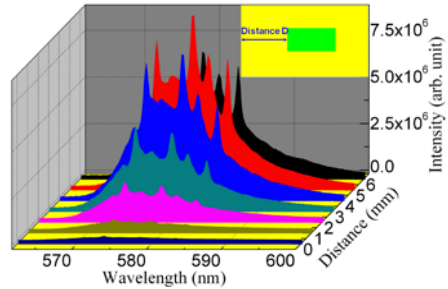

Fig. 3. PL spectra measured with different distances D between the boundaries of the pump beam and the cuvette. The inset at the top right shows the top view of the pumping case. The yellow area is the cuvette. The green zone is the area of the pump beam.

Next, the cross section of the pump beam was fixed as a rectangle (6 mm × 2 mm), and the 6 mm-side was parallel to the detecting direction. The PL spectra were measured with different distances D between the 2-mm side and the boundary of the cuvette. The absorption range of the R6G solution is from 480 nm to 580 nm, meaning that even the dye laser beam can also be reabsorbed. Therefore, the energy loss will be larger than the gain if the feedback is considerably weak. One would expect that, with the increase of the distance D between the 2-mm side of the pump beam and the cuvette boundary, the intensity of the dye laser would decrease while the spectrum is red-shifted due to re-absorption and re-emission. However, the measured results shown in Fig. 3 are obviously different from this expectation. The dye laser becomes stronger with the increase of the distance D, and it reaches maximum strength at a distance of 5 mm. Then, it decreases exponentially with any further increase of D. This phenomenon can be explained by the following mechanism. The size of the equivalent boundary formed by the large gradient of the refractive index in the gain material is larger than the boundary of the pump beam. Thus, the area in which the feedback is provided is larger than the area of the pump beam. When D is about 5 mm, the total feedback is the strongest. Further increase of D means an increase of the equivalent boundary, which leads to a decrease in the gradient of the refractive index. Therefore, the feedback becomes weak and cannot compensate the loss induced by increased re-absorption. As a result, total output energy will decrease. In contrast, when D is less than 5 mm, the smaller equivalent boundary cannot provide enough feedback, so the dye laser cannot reach its maximum energy.

**3. Analysis**

To confirm this new mechanism, we designed the following experiment to study the index change. A He-Ne laser beam (running at 632.8 nm) is incident into the sample from the side of the cuvette as the probe beam to detect the change of the refractivity of the sample. The probe beam is parallel to the surface of the liquid sample, and is about 1 mm below the surface. The pump beam (532 nm) breaks into the liquid gain material R6G from the top of the surface. The boundary of the pump beam is tangent to the probe beam from the top view (as shown in the inset of Fig. 4 (a)). The spot A in Fig. 4 is the spot on the probe beam used to check its direction change after passing through the range close to the pump area in the gain material.

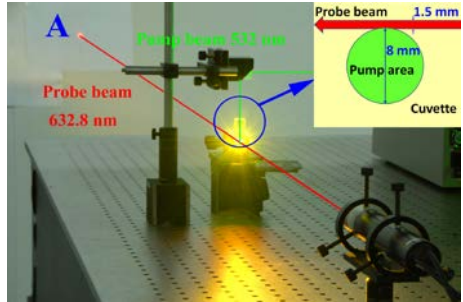

Fig. 4. Optical layout of experiment for checking direction change of probe beam after passing through the pumped gain material. The He-Ne laser (632.8 nm) is incident into the sample from the side of the cuvette as the probe beam. It is parallel to the surface of the liquid sample, and is about 1 mm below the surface. The inset is the top view of the circular area. The boundary of the pump beam is tangent to the probe beam. The spot A is the observed area.

Figure 5 gives the experimental results. Fig. 5(a) shows the probe beam when the R6G is not pumped by the laser. Figure 5(b) shows the probe beam when the R6G is pumped by the laser. When the pump laser is applied, the probe beam passing through the gain material deflects to the left and down. The explanation for this result can be made as follows by intensity-dependent refractive index effect and thermal lensing effects [19].

First, the down deflection of the probe beam is induced by the thermal nonlinear effect: $n = n_0 + \left(\frac{dn}{dT}\right)T_l$, where $n$ is the total refractive index, $n_0$ is the linear refractive index, $\left(\frac{dn}{dT}\right)T_l$ is the nonlinear refractive index, and $T_l$ designates the laser-induced change in temperature. This effect had been used to explain the phenomenon of mirage. In methanol, the refractive index will decrease with the increase of the temperature. In our experiment, the pumping laser of 532 nm is strongly absorbed and less than half of the pumping energy is transferred to heat. The penetration depth of the pump beam is measured to be about 1.7 mm. In other words, most of the heat transferred from the pump beam is concentrated in a very thin layer from the surface of the solution. Hence, there will be a temperature gradient perpendicular to the surface of the solution, which will induce an index gradient ($dn/dz$) in the same direction, but with negative sign. This negative index gradient acts as a negative lens, and turns the probe beam down away from the surface. On another hand, the index gradient along the vertical direction can form a 2-D waveguide parallel to the surface.

Second, however, because the pump area must have a higher temperature than the adjacent area, the horizontal deflection to the pump area of the probe beam should be explained in another way. Due to the convection property of liquid, the temperature gradient along the horizontal direction will be much smaller than that of the vertical direction, so the thermal effect along the surface will also be much weaker, i.e., $\frac{dn}{dz} \Box \frac{dn}{dx}$. We have calculated the isothermal section of the lasing system with a commercial software Comsol. The working time of this system is 7 minutes, and the result is shown in figure 5. It can seen clearly that the temperature gradient along the horizontal direction is much smaller than that of the vertical direction.

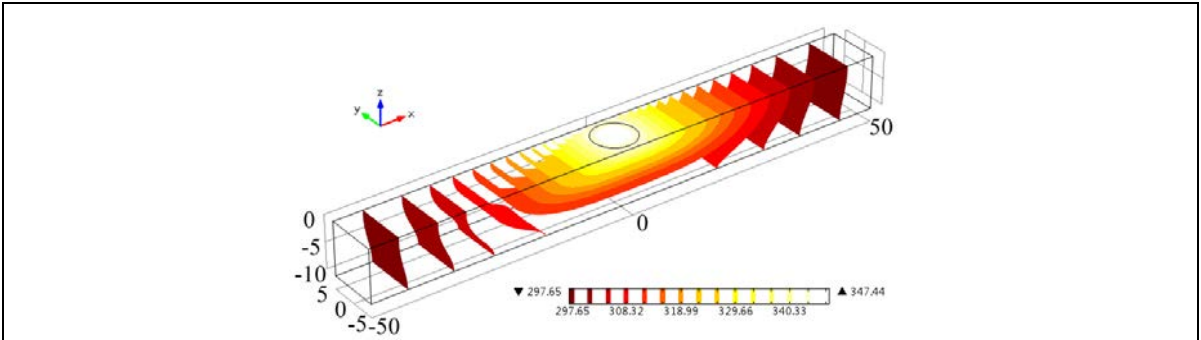

Fig. 5 Stable isothermal section of the laser system.

On the other hand, due to strong resonant absorption, the laser intensity-dependent refractive index effect will play an important role: $n = n_0 + n_2 I$, where $n_2$ is the second order refractive index, $I$ is the laser intensity. In the pump area, this effect will be stronger than the thermal effect along the surface, so the index will be higher than that of the other area. This effect makes the pump area as a "gravitation center" to the probe beam, and form an horizontal equivalent cavity as mentioned above providing effective feedback. From this point of view, the maximum index gradient must be at the boundary of the pump beam. This analysis is in good consistency with the experimental measurement shown in Fig. 6.

According to the above discussions, a thin 2-D wave guide will be formed parallel and close to the surface of the solution. This wave guide together with the equivalent cavity make the lasing emission in transversal direction of the pump beam.

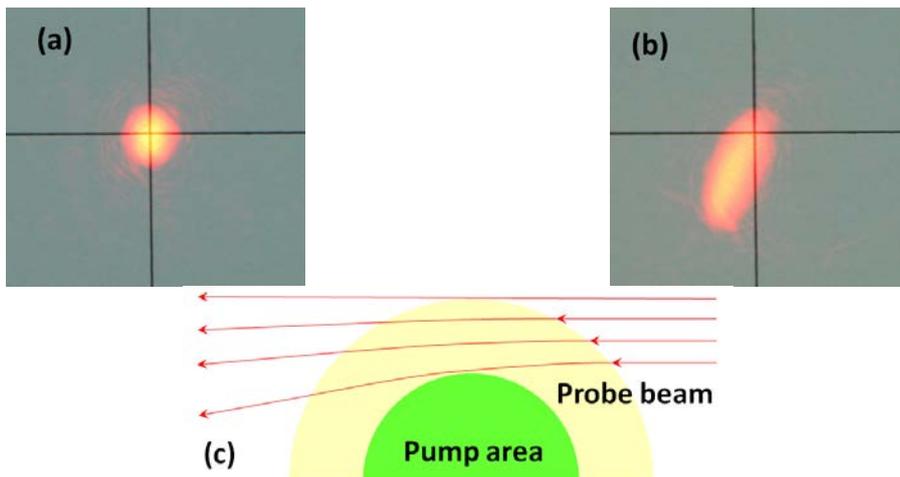

Fig. 6. Experiment results using the optical layout shown in Fig. 4. (a) Spot on the probe beam when the sample is not pumped by the laser. (b) Spot on the probe beam when the sample is pumped by the laser. (c) Schematic demonstration of ray tracing in the range close to the pump area.

## 4. Discussions

It is known that all kinds of lasing are determined by the energy level of the gain material. What is shown in Fig. 7 is the measured 2-D spectrum of R6G in methanol solution. We chose an OPO system (Continuum Sunlite EX OPO) as the pump source during measurements, and changed the pump wavelength from 525 nm to 590 nm. The horizontal axis represents the excited wavelength, while the vertical axis represents the pumping wavelength. Obviously, the diagonal line in Fig. 7 shows the elastic scattering of the pumping laser. This 2-D spectrum can be divided into four areas (See Fig. 7). The areas (a) and (b) correspond to the radiation between the upper laser energy level (the same resonant absorption energy band) and the lower laser energy level (band). This explains why the position of the PL envelope does not change with the pump wavelength. But in area (a), the gain material has stronger absorption, and narrow peaks are observed (the red tiny spots in area (a) of Fig. 7). From Fig. 7 we see that these narrow peaks move synchronously with the wavelength change of the pump beam. Further measurements show that the frequency shifts between the excited peaks and the pumping laser are fixed at 39.5, 41.0 and 45.4 THz. This reveals that the observed phenomenon is the resonant Raman scattering process. The narrow peaks correspond to different sub-energy levels in the absorption energy band, so the positions of the peaks change with the change of the pump wavelength, although the PL envelope does not move. This explains why the peak wavelengths are fixed for different pump conditions (such as different pump power and different distance D as shown in Figs. 2 and 3) at 532 nm. In area (b), the absorption becomes weak; thus, the resonant Raman scattering process does not appear. In area (c), the PL envelope moves with the change of the pump wavelength. Since the pump frequency is away from the absorption band, the resonant absorption condition cannot be satisfied, and the absorption decreases further in area (c). This results in the general Raman scattering in which the PL envelope moves with the change of the pump wavelength. In area (d), the pump frequency is far away from the absorption band, so the absorption is extremely low, and almost no excitation process can occur. It is known that the energy band structure of the dye molecule is very complex. Hence, the spectral property shown in Fig. 7 may also be useful to analyze energy levels and molecular structures of gain materials.

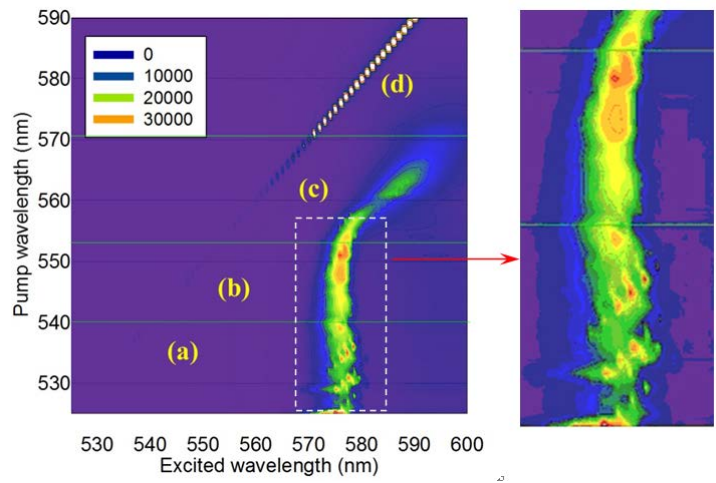

Fig. 7. 2-D PL spectra of R6G solution. The right inset is the enlargement of the part in dashed square.

Figure 8 shows the polarization dependence of the lasing. We measured the spectra of a sample with the same pump power for two orthogonal polarizations. In Fig. 8, the y direction represents the detecting direction parallel to the polarization of the pump beam, while the x direction represents the detecting direction perpendicular to the polarization of the pump beam. Obviously, both the shape and the intensity of the peaks in the x direction are better than those in the y direction. This phenomenon is similar to the polarization property of Rayleigh scattering, although the principle of nonlinear scattering is quite different.

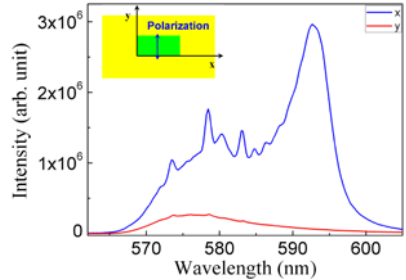

Fig. 8. PL spectra measured for two orthogonal directions x and y perpendicular to and parallel to the polarization of the pump beam, respectively. The inset at top left shows the top view of the pumping case. The yellow area is the cuvette. The green zone is the area of the pump beam.

Figure 9(a) shows a picture of lasing in the system. Obviously, the excited radiation is almost parallel to the surface of the sample. This means that the dye laser beam can be easily coupled into a waveguide and form special output modes. All the radiation directions in Fig. 9(a) can be calculated by the coupling of the waveguide modes. It can be seen clearly that the divergence of our 2D laser is very small. This provides a significant potential application in integrated optics. We fabricated several similar laser systems with different shapes. The thin silica plate was used as the substrate. A groove with different shapes was dug at the central part of the substrate. The groove was filled with liquid gain material, and the substrate played the role of waveguide. The pump beam was vertically incident on the gain material in the groove from the top. To obtain uniform output radiation, the pump beam should be circularly polarized based on the discussion about Fig. 8. The excited radiation will be coupled into the waveguide automatically. Figure 9(b) shows one sample with the shape of circle. It can be seen that this disk laser achieves 2-D omnidirectional radiation, and the radiation in all directions is uniform. This property will be useful in optical communications and integrated optics, such as wavelength-division multiplexing and signal sharing. The total energy conversion efficiency is measured to be over 28%, which is even higher than the commercial dye laser. Figure 10 shows the temporal profiles of the pumping laser and the excited radiation. It can be seen that this disk laser maintains a good temporal profile, and its pulse duration is almost the same as the pump duration. It is important in optical information processing when using nonlinear method without signal distortion.

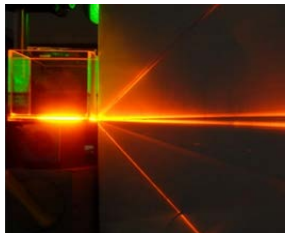   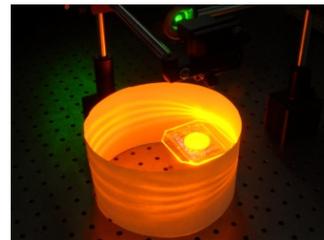

(a) Output of the dye laser beam.        (b) Multi modes radiation with circular groove.
Fig. 9 Output character of liquid disk laser system.

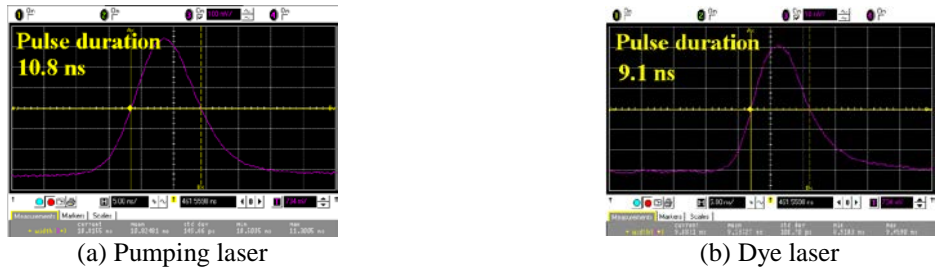

(a) Pumping laser  (b) Dye laser

Fig. 10 Temporal profiles of the pumping laser and the dye laser.

**Acknowledgements**

The authors thank the National Natural Science Foundation of China (Grant No. 11074024, 11104016, 10904003) for financial support. Also, the authors would like to thank Prof. Jing Zhou, Dr. Tianrui Zhai and Dr. Zhaona Wang for helpful discussions.